\documentclass[12pt,preprint]{aastex}

\newcommand{\etal}{et~al.}

\slugcomment{To be published in ApJ, 2009} 
\shorttitle{Sextans Dwarf Spheroidal Galaxy}
\shortauthors{Lee et al.}

\begin{document}

\title{Star Formation History and Chemical Evolution 
of the Sextans Dwarf Spheroidal Galaxy}

\author{
Myung Gyoon Lee\altaffilmark{1},
In-Soo Yuk\altaffilmark{1,2},
Hong Soo Park\altaffilmark{1},
Jason Harris\altaffilmark{3},
Dennis Zaritsky\altaffilmark{4}
}

\email{
mglee@astro.snu.ac.kr;
yukis@kasi.re.kr; hspark@astro.snu.ac.kr;
lmcboy@gmail.com; 
dzaritsky@as.arizona.edu}

\altaffiltext{1}{Astronomy Program, Department of Physics and Astronomy, 
Seoul National University, 56-1 Sillim 9-dong, Gwanak-gu, Seoul 151-742, Korea}

\altaffiltext{2}{Korea Astronomy and Space Science Institute,
61-1 Hwaam-dong, Yuseong-gu, Daejeon, 305-348, Korea}

\altaffiltext{3}{Space Telescope Science Institute,
3700 San Martin Drive, Baltimore, MD 21218}

\altaffiltext{4}{Steward Observatory, University of Arizona,
933 North Cherry Avenue, Tucson, AZ 85721-0065}

\begin{abstract}

We present the star formation history and chemical evolution 
of the Sextans dSph dwarf galaxy as a function of
galactocentric distance. We derive these from the $VI$
photometry of stars in the $42' \times 28'$ field
using the SMART model developed by Yuk \& Lee (2007, ApJ, 668, 876)
and adopting a closed-box model for chemical evolution.
For the adopted age of Sextans 15 Gyr,
we find that $>$84\% of the stars formed prior to 11 Gyr ago, significant star formation
extends from 15 to 11 Gyr ago ($\sim$ 65\% of the stars formed
13 to 15 Gyr ago while $\sim$ 25\% formed 11 to 13 Gyr ago), detectable star formation 
continued to at least 8 Gyr ago, the star formation history is
more extended in the central regions than the outskirts, and the difference in star formation rates
between the central and outer regions is most marked 11 to 13 Gyr ago.
Whether blue straggler stars are interpreted as intermediate age 
main sequence stars affects conclusions regarding the star formation
history for times 4 to 8 Gyr ago, but this is at most only a trace population.
We find that the metallicity of the stars increased rapidly up to [Fe/H]=--1.6  
in the central region and to [Fe/H]=--1.8 in the outer region 
within the first Gyr, and has varied slowly since then.  
The abundance ratios of several elements derived in this study
are in good agreement with the observational data based on the high
resolution spectroscopy in the literature.
We conclude that the primary driver for the radial gradient 
of the stellar population in this galaxy is the star formation history, which 
self-consistently drives the chemical enrichment history. 
\end{abstract}

\keywords{galaxies:stellar content --- galaxies:chemical
abundance---galaxies:evolution --- galaxies: individual(Sextans)}

\section{Introduction}

We know of about a dozen dwarf spheroidal (dSph) satellite galaxies of 
the Milky Way Galaxy \citep{mat98, bel06,tol09}. Although these galaxies
are among the faintest and least massive, they 
are, judging from the discovered population in the Local Group, 
the most common  in the Universe. 
As such, they provide interesting constraints on models of galaxy
formation and evolution in a cosmological context. The nearest of these, 
because of the increased sensitivity and angular resolution,
provide the optimal environment in which to investigate
the star formation history (SFH) of low-mass galaxies. In turn, such studies
inform investigations of the first stars in the Universe 
\citep{wys05,ric05,kaw06, fen06,bro09} and the 
stellar halo of the Milky Way Galaxy 

For various reasons,  the SFH of  these galaxies was originally presumed to 
consist of a simple single burst of star formation at early times.
In part, this was the natural extension of the known characteristics of previously identified
halo populations (the old
stellar halo and the old, single burst population of globular clusters).
However, deep photometric studies found that the SFH of dSph's is much more complex and 
intriguing than expected \citep[e.g,][]{lee93, mat98,dol02, dol05, tol09} and have
ultimately shown that simple star formation histories are the exception rather than the rule.

The Sextans dSph, discovered in 1990 \citep{irw90}, 
is one of the fainter known galaxies ($M_V>-12$ mag) in the Local Group.
\citet{lee03} presented deep, wide field  $VI$ photometry of this galaxy
 based on the CFH12K images, and discussed inferences drawn from the color-magnitude diagram.
To summarize, \citet{lee03} derived 
a schematic star formation history of the Sextans dSph and found the following:
1)  most stars in Sextans formed at a similar time, and with similar low metallicity, 
as the metal-poor globular, M92,
2) hints of continuing star  formation 
in the inferred metallicity spread ($\Delta$[Fe/H]$\approx 0.2$ dex) among
red giant branch stars, population gradients, strong red horizontal branch, 
and bright main-sequence stars
slightly above the main sequence turnoff of the main population (see also \citet{iku02})
and 3) a small fraction of all the stars, including anomalous Cepheids, formed less than 1 Gyr ago.

Somewhat in contrast,  the large  metallicity range of the individual stars 
($-1.45 \le  $[Fe/H]$ \le -2.85$),
derived for five red giant stars in Sextans by
\citet{she01} from high resolution spectra obtained with 
the High Resolution Echelle Spectrometer (HIRES)
on the Keck I telescope is inconsistent with the hypothesis of a dominant, short-lived
star formation episode.
The existence of such a metallicity spread differs from what is observed in globular clusters and requires
some self-enrichment, which in turn implies an extended SFH.

To resolve the potential discrepancy among various studies, we reanalyze the \citet{lee03} photometry
using the SMART model, a new numerical algorithm developed by \citet{yuk07}, which
quantitatively recovers the 
SFH and chemical evolution
by synthesizing the color-magnitude diagram. 
We discuss the following as follows:
in \S 2 the SMART model used to derive the SFH
and chemical evolution, and the available data for Sextans,  in
\S 3 the results of the modeling, in
\S 4 comparisons to other models and the implications of the results, 
and in \S5 our summary of the main results.

\section{Model and Data}

\subsection{Model}

The SMART model recovers the 
SFH and chemical evolution of a stellar system 
by synthesizing the color-magnitude diagram in comparison with observations.  
The principal difference between 
the numerical algorithm in SMART and other algorithms
used to derive the SFH of resolved stellar populations is that SMART
treats the internal chemical evolution self-consistently.
The detailed description of this model is given in \citet{yuk07}, 
and so we describe it briefly here.
The core aim of this model is to simultaneously derive 
the SFH and chemical evolution of galaxies from observational color-magnitude diagrams (CMDs).
The algorithm  produces a set of simulated CMDs for varying
star formation histories, and compares these to 
the observed CMD to find the best-fitting SFH.
In this model we use the Padova database of theoretical
isochrones\citep{gir02}. The Padova isochrones cover the evolution
of stars from zero-age main sequence to carbon ignition for
higher mass stars and to the thermally pulsating asymptotic giant
branch phase for lower mass stars. We preform interpolation to make a fine grid
using logarithmic variables according to the evolutionary equivalent
point of evolutionary tracks.

The best-fit 
is determined by minimizing the goodness-of-fit parameter $\chi^2$.
If the best-fit star formation rate at the $k$-th time bin
is $\phi_k^{sol}$, then the asymmetric confidence limits of the
$k$-th time bin can be obtained as follows. We find the best-fit
star formation rate with the fixed $\phi_k^{sol} + \delta \phi_k$,
and calculate goodness-of-fit parameter  $\chi^2_p (\phi_k^{sol} +
\delta \phi_k)$. 
If the difference between $\chi^2_p (\phi_k^{sol})$
and $\chi^2_p (\phi_k^{sol} + \delta \phi_k)$ is $1\sigma$, then
$\delta \phi_k$ is the $1\sigma$ confidence limit of the $k$-th time
bin. 
In the similar way, we calculate confidence limits of each 
time bin in plus and minus direction.

\subsection{Data}

We use the $VI$ photometry of stars in a $28'  \times 42'$ field centered on Sextans given by \citet{lee03}. This photometric catalog comes from
CCD images obtained using the CFH12K camera at the CFHT.
The limiting magnitudes (50\% completeness) are $V = 24.4$ mag and $I = 23.6$ mag.
To investigate spatial variations in the SFH, 
\citet{lee03} radially divided the entire field into four different regions:
R1($<7\arcmin.3$), R2($7\arcmin.3-11\arcmin$), R3($11\arcmin-15\arcmin$), and R4($>15\arcmin$).
The boundaries of the regions are set to maintain a constant number of
mid-red giant branch (RGB) stars with $0.8 < (V-I) < 1.1$ mag and $18 < V < 21$ mag. The properties of these regions are summarized in Table 1. 
For reference, the boundary of region R3 (15 arcmin) is slightly smaller 
than the core radius of the Sextans dSph $16\arcmin.6 \pm 1\arcmin.2$ \citep{irw95}.

\begin{figure} 
\epsscale{.80}
\plotone{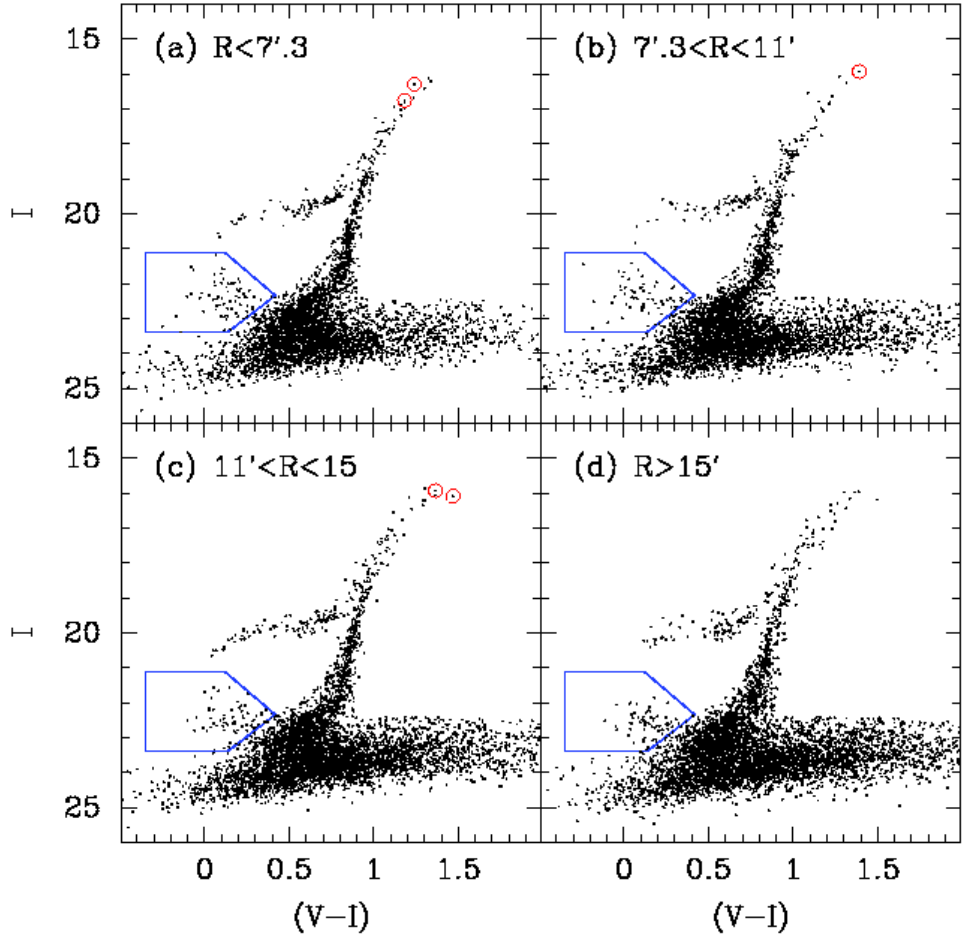} 
\caption
{$I-(V-I)$ diagrams of the four regions at different radii from the center of the
Sextans dSph.
Open circles represent the stars with high resolution spectroscopic data
given by \citet{she01}.
Pentagons represent the region of the blue straggler stars.
\label{cmd_region}}
\end{figure}

In Figure~\ref{cmd_region} we present the observed stellar CMDs 
in each of the four regions of the Sextans dSph given by \citet{lee03}.
Bright foreground stars 
were not used in modeling. 
The five red giant branch stars for which chemical abundances have been measured \citet{she01}
are marked with circles.
Basic information for these giant stars is presented in Table 2.

One outstanding feature of this galaxy is its large number of blue stragglers
\citep{lee03}, which we highlight with pentagons in Figure 1.
There are  59, 64, 68, and 69 such stars
in R1, R2, R3, and R4, respectively.

\section{Results}

We run the SMART algorithm in each of the four regions independently
to investigate whether there are radial SFH variations.
The model parameters we adopt  for our calculation are summarized in Table~\ref{model_par}
and we comment on a few of our choices.
Previous estimates for the distance to Sextans range from $(m-M)_0 = 19.7$ to 19.9, 
and the reddening values range from $E(B-V)=0.01$ to 0.03 (see the summary in \citet{lee03}).
Comparing the observational CMD and the synthetic CMD, we determined which distance produces
the best model fits, obtaining 
a distance modulus $(m-M)_0 = 19.67$ and reddening $E(B-V) = 0.03$
($E(V-I) = 0.04$).
This value of the distance is slightly smaller than that derived in \citet{lee03}.
At this distance one arcmin corresponds to 25.0 pc.
Our choice of Salpeter's initial mass function ($\phi(m)=A m^{x-1}$ ) with
an index of $x=-1.35$ \citep{sal55}, is supported by 
\citet{wys05} who found
that the initial mass function in Ursa Minor,
a dSph similar to Sextans, is consistent this function.
We adopted a closed-box model for chemical evolution model.
While \citet{fen06} and \citet{iku02} suggested in the study of chemical evolution of dSphs that the galactic winds
do not make a major role in removing gas from dSphs, 
\citet{lan04, lan07} and \citet{rec08} presented a different view: chemical properties of the
dSphs can be explained well only by the chemical evolution  model involving
intense galactic winds and low star formation efficiency.
In this study we will show that the closed box model also reproduces well
the observed abundance ratios in Sextans. 
We also assume that metals produced in a region stay in that region
(i.e., very inefficient mixing).
We used the $VI$ photometry brighter than the 60\% completeness limit
for modeling. 
We also assume that  the current stellar mass of the galaxy was the same as 
the gas mass in the beginning.

\citet{tol03} measured the ages of red giants with known abundances in this galaxy to
be 15 Gyr using stellar models. This age is larger than the age of the Universe derived from current high-precision cosmological models \citep{kom09} and suggests that there is an unresolved problem with the stellar models. 
Matching the stellar ages of old populations with the age of the Universe has been a long-standing problem, and could even be considered the first evidence for a cosmological constant \citep{bol95}; 
however, the current level of discrepancy is within the range of various plausible model adjustments. We adopt an age of 15 Gyr so that the current stellar models match the oldest populations, but caution against an absolute interpretation of the derived ages and favor a relative interpretation. 
We used seven age bins for calculation: 0--4, 4--7, 7--9, 9--11, 11--13, 13--14, and  14--15 Gyr.

\subsection{Star Formation History}

\begin{figure}  
\epsscale{.50}
\plotone{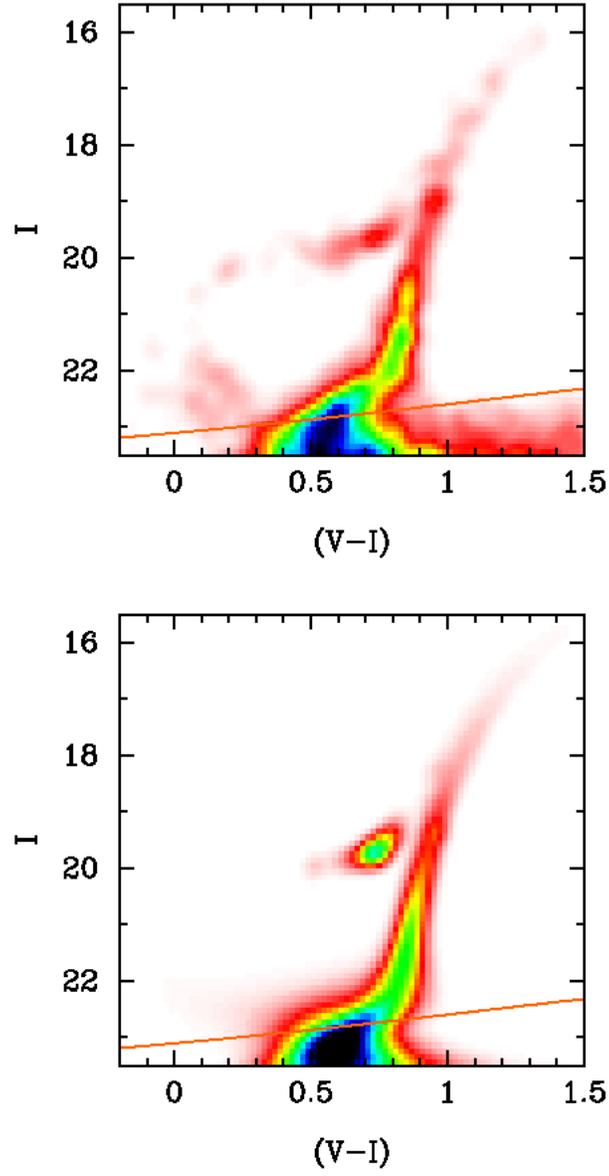} 
\caption
{The observational CMD (upper panel) and the best fit synthetic CMD for Case A (lower panel) 
of the Sextans dSph.
The solid line corresponds to the 60\% completeness level of the photometry.
\label{cmdos}}
\end{figure}

We derive the SFHs of each of the four regions 
for the two blue-straggler scenarios (Case A and Case B). 
We normalized the star formation rate with respect to the lifetime average star formation rate for the entire observed region of Sextans, 
$1.7 \times 10^{-4}$ $M_\odot$ yr$^{-1}$. 
In Figure~\ref{cmdos} we present the observed CMD for the entire observed area and
the best-fit synthetic CMD for Case A.
The two CMDs look very similar in general, showing that the synthetic CMD reproduces 
well the observed CMD.
We describe the resulting SFHs for Case A and Case B below. 

\subsubsection{Case A : Including Blue Straggler Stars}

In Table~\ref{sfh_w_bs_tbl} we list the best-fit, Case A SFH and in Figure 3
we display it.
The star formation rates are weighted by the number density of each region,
and the errors correspond to the 1 $\sigma$ confidence level.
From the global SFH shown in Figure 3, we conclude 
that star formation was extended, and not a single burst like
in Galactic globular clusters.
While most star formation occurred between 15 and 7 Gyr ago, detectable
star formation continued until 4 Gyr ago.
To be more quantitative, we find that
57 \% of the stars formed between 13 and 15 Gyr ago, 
27 \% of stars formed between 11 and 13 Gyr ago, 
the remainder formed mostly until 7 Gyr ago, and a tracer population, which
consists of the blue stragglers, formed until 4 Gyr ago.
Thus, a majority (84 \%) of the stars formed within the first 4 Gyr.

The star formation rate decreased more rapidly in the outer regions.
In the central region (R1), 52 \%, 32 \%, and 10 \% of stars formed
$13-15$, $11-13$, and $9-11$ Gyr ago, respectively, while
in the outer region (R4), 72 \%, 15 \%, and 8 \% of stars formed
during the same periods.
Thus, the star formation rate in the central region decreased by a factor of five
from 14 Gyr to 10 Gyr ago, 
while that in the outer region decreased by a factor of nine.
The most distinct period occurred 11--13 Gyr ago, when the star formation rate
in the inner region was a factor of two larger than in the outer region.

In Figure~\ref{sfh_rad_w_bs} we display the radial  behavior of the
star formation rate within each age bin.
There is an evident radial gradient in the star formation rate at all times when the
star formation is significantly different than zero.
Prior to 11 Gyr ago, the star formation rate decreases approximately linearly with increasing radial distance,
and the gradient is steepest during the period 11--13 Gyr ago.
Therefore, it is this population of stars, formed between 11 and 13 Gyr ago, that is 
primarily responsible for the observed population gradient.

\subsubsection{Case B : Excluding Blue Straggler Stars}

The Case B SFH of each region is listed
in Table~\ref{sfh_wo_bs_tbl} and is displayed in Figure~\ref{sfh_wo_bs}.
While the pattern of star formation is similar to the Case A results, 
one notable difference arises due to the exclusion of blue straggler stars in Case B:
the Case B, SFH does not contain the tracer population that forms between 4--7 Gyr.

Again, we find that the star formation decreases more rapidly in the outer regions.
In the central region, R1, 56 \%, 34 \%, and 7\% of stars formed
$13-15$, $11-13$, and $9-11$ Gyr ago, respectively, while
in the outer region, R4,
81 \%, 11 \%, and 6 \% of stars formed
over the same periods. 
In total, we find that 
65 \% of the stars formed between 13 and 15 Gyr ago, 
25 \% of stars were formed between 11 and 13 Gyr ago, 
and the remainder formed until 7 Gyr ago.  Again, we find that the
vast majority (90\%) of stars formed prior to 11 Gyr ago. This quantitative measurement
varies only slightly between Case A (84\%) and Case B (90\%).

In Figure~\ref{sfh_rad_wo_bs} we display
the radial variation of the star formation rate derived for Case B.
As in Case A,
the radial gradient of the star formation rate is prominent in age bins where we
measure a non-zero star formation rate.
The star formation rate decreases approximately linearly with increasing radial distance 
from the beginning, 
and the radial gradient is steepest 11--13 Gyr ago.
The shape of the radial distribution appears almost identical to that for Case A.

\subsection{Chemical Evolution: Metal Enrichment History}

The results for chemical evolution are very similar between Cases A and B
so that we describe only the results for Case B hereafter.
In Figure~\ref{feh_wo_bs} we present the metal enrichment histories of each of the
four regions in the Sextans dSph for Case B.
We plot the mass-weighted mean metallicity of the stars versus time.
The metallicity increased rapidly within $\sim 0.1$ Gyr, and more slowly thereafter.

Radially, the metallicity of the stars increased to [Fe/H]=--1.6  
in the central region and to [Fe/H]=--1.8 in the outer region 
within the first Gyr. 
The early metal enrichment (within 2 Gyr from the beginning) of R1 coincides with that of R2, 
because the star formation rate in R1 is nearly the same as that in R2.
The metallicity difference between R1 and R4 was
$\Delta\rm{[Fe/H]} \approx 0.15$ until 14 Gyr ago, 
and $ \approx 0.33$ 7 Gyr ago.

In Figure~\ref{feh_rad_wo_bs} we display the radial metallicity distributions 
for four different ages (14.98, 14.9, 14.0, and 7 Gyr).
The radial gradient is zero within $R = 10'$ and is --0.017 dex/arcmin at $R > 10'$
within the first Gyr. 
The slope of the radial gradient becomes constant 
over the entire region ($\sim -0.03$ dex/arcmin) 7 Gyr ago.

Previous estimates of the mean metallicity of stars in the Sextans dSph, as summarized by \citet{lee03},
were derived from the photometry of a large number
of red giants, low-resolution spectroscopy, or high-resolution spectroscopy of a
small number of red giants.
Interestingly, the distribution of these estimates 
is roughly bimodal,
with one peak at [Fe/H]$\sim-1.6$ dex \citep{mat91, dac91, mat95, mat98},
and the other at [Fe/H]$\sim -2.05 $ dex \citep{sun93,gei96,she01}.
The bimodality is present 
among both the photometric and spectroscopic estimates.
No reason for this bimodal distribution is known.
\citet{lee03} concluded on the basis of averaging 
that the mean metallicity of the giants in the Sextans dSphs
is $\langle [Fe/H] \rangle =-2.1\pm0.2 $ with a dispersion of 0.2 dex. 
The metallicity estimates for five red giants based on high-resolution spectra 
show a large range from [Fe/H]=--2.85 to --1.45, with a mean of $-2.07 \pm 0.21$ \citep{she01}. However, this is based on only five stars so we
need to increase the sample size.
 
In Table~\ref{feh_mean_tbl} we present the mean metallicities from our model.
Our derived values of  $\langle$[Fe/H]$\rangle$ are independent of whether we include
or exclude blue straggler stars. For our entire Sextans survey region, we find that 
$\langle$[Fe/H]$\rangle = -1.76$ dex, 
which is 0.3 dex larger than the values given by \citet{she01} and \citet{lee03}. However, 
this difference is insignificant ($< 2\sigma$) because of the large dispersion among 
the observed [Fe/H] for individual stars and the small number of stars with spectroscopic 
measurements.

\subsection{Chemical Evolution: Chemical Abundance Ratios}

The evolution of the chemical abundance of a galaxy depends on
its SFH because 
different elements are produced during the evolution of stars with different masses
over a range of timescales \citep{mat08}.
A nice summary of the main features of the chemical evolutionary history of
different elements is given by \citet{tol03}.
Most elements used in the study of chemical evolution of stellar populations belong
to four groups:  1) light elements including O, Na, Mg, and Al,
2) $\alpha$-elements (or even-Z element) including O, Mg, Si, Ca, and Ti,
3) Iron-peak elements including V, Cr, Mn, Co, Ni, Cu, and Zn,
and 4) heavy elements including Y, Ba, Ce, Sm, and Eu.
Light elements are a tracer of deep-mixing abundance patterns in RGB stars, 
and they show significantly different patterns
between globular cluster giants and field stars.
$\alpha$-elements are mostly produced in massive stars and ejected in SNe II.
Iron-peak elements are produced via the explosive nucleosynthesis sequence during the nuclear
fusion (in both SN Ia and SNe II). 
Cu and Mn are considered to be produced by SNe Ia, in which
case the ratio [Cu/$\alpha$] or [Mn/$\alpha$] is a good indicator of the ratio between SNe Ia
and SNe II.
Heavy elements provide a measure of
the ratio of $s$-process (occurring in AGB stars) to $r$-process (occurring in SNe II)
elements.

A particularly informative parameter in the study of chemical evolution is the ratio [O/Fe].
Oxygen is produced primarily in high-mass stars of very short lifetimes
and is ejected by SNe II, while iron is produced in both SNe II and SNe Ia.
In Figure~\ref{ofe_age_w_bs} 
we show the variation of [O/Fe] over time derived from 
Case B model.
The value of [O/Fe] decreases very rapidly in the beginning, then more
slowly after reaching 
[O/Fe]$ \approx 0.08$,  and then even more slowly after 5 Gyr. 
[O/Fe] is similar to the solar value at about 11 Gyr ago.
Radially, [O/Fe] varies little, except for 
12 to 9 Gyr ago when it was slightly higher in the inner region than
in the outer region.  
Unfortunately, it is difficult to observationally determine [O/Fe] for stars,
and the value of [O/H] is typically derived using emission line regions in a galaxy.
Therefore, there are no available stellar [O/Fe] measurements in Sextans 
that can be compared to our predictions.

Another informative parameter is the ratio [$\alpha$/Fe].
This ratio, like [O/Fe], depends on the relative contribution of SNe II and SN Ia. Unlike
[O/Fe] it can be measured from stellar spectra.
The timescale for changes in [$\alpha$/Fe] depends not only on the SFH,
but also on other factors such as the initial mass function,
the SNe Ia timescale, and the timescales for mixing the SNe Ia and SNe II products
back into the interstellar medium \citep{mat03, mat03b}.
We use a definition of the average [$\alpha$/Fe] as an average of Mg, Ca, and Ti abundances
($\rm{ [\alpha/Fe] = [{1 \over 3} [(Mg + Ca + Ti)/Fe] }$) 
following \citet{she01} and \cite{tol03}.
In Figure~\ref{alpha_age_w_bs} 
we plot  [$\alpha$/Fe] vs. time for our Case B model 
and the observed values from five red giants in Sextans 
\citep{she01, tol03}. 

The evolution of [$\alpha$/Fe] is very similar to that of [O/Fe].
Our models are generally consistent with the observations,
but differ somewhat from the observations of the star S58.
While our model predicts [$\alpha$/Fe] $\approx -0.1$ dex since 9 Gyr ago, 
the observed value for this one star is [$\alpha$/Fe]$\approx -0.26$. 
To fit the star S58, [$\alpha$/Fe] should decrease abruptly as [Fe/H] increases.
We tried to estimate the age of S58 and the other four stars 
using the new photometry given by \citet{lee03}
and the revised Yonsei-Yale isochrones \citep{dem04} in the similar way to that used
in \citet{tol03}. The estimated values for S58 are 7 Gyr from $(B-V)$ color, and
6 Gyr from $(V-I)$ color, which are similar to the value given by \citet{tol03}.
The ages derived for the other stars are larger than 14 Gyr, similar to those given
by \citet{tol03}.

In Figure~\ref{alpha_w_bs} we plot [$\alpha$/Fe] versus [Fe/H] for the four radial regions in Sextans 
for Case B. 
[$\alpha$/Fe] changes slowly about the value of 0.1, but then decreases sharply
for [Fe/H]$ >  -2$. The predictions reach the solar value at [Fe/H]$\approx -1.7$,
and continue to drop to the 
the minimum value of $\sim$ --0.1.
The results from our modeling are consistent with the observations of four stars, 
but again differ for the star S58.

Abundance ratios like [O/Fe] and [$\alpha$/Fe] are affected by the rates of SN Ia and SN II
during the evolution of a galaxy.
In general SN II's have a significant effect from the beginning of the star formation history in
galaxies with strong initial starbursts, while SN Ia's begin to play an important role in setting the abundance ratios after
the time when the rate of SN Ia is a maximum,  $t_{SN Ia}$
(the typical time scale for the SN Ia maximum enrichment) \citep{tim95,mat01}.
The value of $t_{SN Ia}$ is as small as $0.3 - 0.5$ Gyr for elliptical galaxies and 
the bulges of spiral galaxies,
in which the star formation rate was high in the beginning,  while it
is larger, $1.5$ Gyr, for our Galaxy, and can be 
as large as $7 - 8$ Gyr for irregular galaxies,
in which stars formed over a much more extended period of time \citep{mat01}. 
In Figure~\ref{snr_w_bs}
we show the evolution of the SN II and SN Ia rates for our Case B model. 
The SN II rate was the maximum in the beginning, 
decreasing thereafter until the end of star formation.
The SN Ia rate increased slowly from zero since the beginning, 
reaching the maximum  13--11 Gyr ago, when the star formation was active in the inner regions, 
and decreasing thereafter.
Thus, $t_{SN Ia}$ is  2 to 4 Gyr for the Sextans dSph.
This value is much larger than that for elliptical galaxies,  but 
is between those of our Galaxy and irregular galaxies. 

Other abundance patterns,   [X/Fe] vs. [Fe/H] , 
can also be useful in constraining the star formation and the
chemical enrichment history \citep{tol03}.
In Figures~\ref{ratio_wo_bs} 
we display the abundance
ratio of several elements for our Case B model. 
The observations presented by \citet{she01} are plotted for comparison.
The model abundance ratios are generally in good agreement with the observational ones.
It is noted that the model values for {\rm [V/Fe]} and {\rm [Cr/Fe]} are marginally larger and smaller than the observation values, respectively, but the differences
are at the level of one sigma.

In Figures ~\ref{isotopes_wo_bs}, we 
show the abundance of present stable isotopes
normalized with respect to the solar abundance \citep{and89} for our
Case B model. 
The ratios of most of the heavy elements are between 0.1 to 0.001.

\section{Discussion}

\subsection{Comparison to other models}

\citet{iku02} investigated the star formation and chemical enrichment history of
three dSphs (Draco, Sextans, and Ursa Minor) using a chemical evolution model and
a simulation program for CMDs. 
Like us, they adopted a Salpeter IMF and a closed box model.
They concluded that the observed abundance of the giants and the CMD of the stars
in Draco could be explained by very low star formation rates that continued for a long time since 
the beginning (3.9--6.5 Gyr).
When they adopted much steeper IMFs ($x=-1.75$ to --2.15) and used a star formation rate similar to that of the solar neighborhood, they found that the star formation time could be shorter, 1.6--2.2. Gyr.
They did not  model Sextans, but noted that Sextan's SFH 
might be similar to that of Draco because of 
the analogous features in the CMD, a narrow RGB and red HB.
Our result, that star formation continued for 8 Gyr in Sextans, is consistent with their
inference based on their modeling of Draco with the same IMF.

In contrast to our results, 
the star formation rate in the \citet{iku02} models does not decrease rapidly 
because it depends on the gas density, which declines 
slowly due to the very low star formation rates. 
Hence, the resulting SFH should differ qualitatively from ours.
Such differences will be reflected in the evolution of  [Fe/H]. 
In \citet{iku02} (see B and C models in their Figure 8), [Fe/H] increases slowly from the beginning,
reaching [Fe/H] = --2.4 to --2.7 dex at 1 Gyr, 
and then increases faster reaching [Fe/H] = --1.4, a few Gyr later.
In our model, metallicity
increases rapidly during the first 1 Gyr, but then changes little
(Figures~\ref{feh_wo_bs}). 
This behavior is 
consistent with the results for the first galaxies (similar to the Local Group dSphs) 
given by \citet{kaw06}.
The differences between the results presented here and those of 
\citet{iku02} are  due to difference in the star formation rate.

\citet{lan04} studied the chemical evolution of Sextans and other dwarf spheroidal galaxies using the outflow model of chemical evolution, 
and found that their best model to match the observed abundance ratios of these dSphs
is a combination of a high wind efficiency and a very low star formation 
efficiency. Their modeling results for [Si/Fe], [Mg/Fe], and [Ca/Fe] in the case of Sextans are consistent with the observed values (in their Figure 4), like ours.
It is interesting that both our models based on a closed-box model
and theirs based on a high galactic wind model are consistent with the
observed abundance ratio. 
It is noted that we derived the star formation rate directly from the
color-magnitude diagram, while they inferred it from the color-magnitude diagrams.

\subsection{Radial Gradients of Stellar Populations in dSphs}

Younger stellar populations with higher metallicity are often
concentrated toward the centers of dSphs, a phenomenon 
often referred to as radial population gradients.
For example, red horizontal branch stars are more centrally concentrated than
blue horizontal branch stars \citep{harb01,lee03}.
\citet{harb01} discussed several possible explanations of
these population gradients 
and concluded that metallicity is likely to be a main driver
in some galaxies, including Sextans, while age is 
the critical factor in others, such as Carina.
\citet{riz04} investigated the spatial distribution of the SFH
of Carina, Sculptor, and Sextans and concluded that
the driver was age for Carina, metallicity for Sculptor, and both of age and metallicity for
Sextans.
All of these require some spread in the characteristics of the stellar populations and
both age and metallicity have been invoked for Sextans.
\citet{bel01} argued that Sextans  has two old populations
of different metallicity, [Fe/H]=--2.3 and --1.8, but there is no other observational evidence for
such a population mix \citep{lee03}.
Lastly, \citet{lee03} concluded that the wide color distribution for
the upper red giant branch is caused by continuous star formation, and that the metallicity
increases during the period of a few Gyr, leading to $\Delta$[Fe/H]$\approx 0.2$ dex.

Our results support the findings of \citet{lee03} and \citet{riz04}.
The spatial variation of the star formation rate is obvious in this study:
the star formation rate is higher in the central region than in the outer region.
As star formation proceeds, the metallicity increases with time. This
results in the radial gradient of metallicity, which is consistent with the radial
gradient of the stellar population in Sextans \citep{harb01,lee03}.
We conclude that the primary driver for the population gradient in Sextans
is the SFH. 
It does not make any sense to argue for only either age
or metallicity being the cause since they are related.

\subsection{Blue Straggler Stars}

If blue straggler stars are intermediate age main sequence stars,
then they formed between $\sim 4$ and $\sim 7$ Gyr ago.
\citet{lee03} pointed out, that in this case
one expects red giant clump stars to be present between
the red and blue horizontal branches, which are not seen
in the observed CMD.
Therefore, it is unlikely that these blue stragglers are
indeed intermediate-age normal main sequence stars.
Similarly, \citet{car02} concluded that the Ursa Minor blue straggler populations 
originates in the old population.
On the other hand, \citet{map09} studied the spatial distribution of blue straggler stars based on wide field imaging, concluding  
that most of the blue straggler stars in the
Fornax dSph are young main sequence stars, while the blue straggler stars in the Sculptor dSph are old binaries with mass-transfer.

Our model cannot clarify the nature of the blue straggler stars, but can illustrate differences
arising in the evolution of Sextans in the two scenarios. As we have shown,
the SFH from 4 to 7 Gyr ago is where changes occur.
However, the effect of this change on the chemical evolution is negligible
because the number of the blue straggler stars is small in comparison to the
rest of the population and the bulk of stars formed prior to this time.

\section{Summary and Conclusion}

Using the SMART model developed by Yuk \& Lee (2007), 
we derive the SFH and chemical evolution of 
the Sextans dSph from the $VI$
photometry of stars over the $42' \times 28'$ field described
by \citet{lee03}. Our principals results are summarized below.

\begin{enumerate}

\item  
Independent of our assumption regarding the nature of blue straggler stars,
the principal star formation episode continued for several Gyr.
From models in which we exclude the blue stragglers,  we find that 
about 65 \% of the stars formed within the first two Gyr,
about 25 \% of stars formed between 13 Gyr ago and 11
Gyr ago, and the bulk of the remainder formed up until 7 Gyr ago. 
The models in which we include the blue stragglers and treat them as intermediate
age main sequence stars,  differ principally in that
star formation continues at a low level until 4 Gyr ago.
While the effect of the blue straggler stars on the SFH
between 7 and 4 Gyr ago is noticeable, 
their effect on the chemical evolution is negligible.

\item
The mean metallicity of the stars increases rapidly up to [Fe/H]=--1.6  
in the central region and [Fe/H]=--1.8 in the outer region for stars created within the first Gyr, 
and varies slowly for younger stars.

\item
Our derived abundance ratios of several elements  are generally 
in good agreement with measurements based on high
resolution spectroscopy \citep{she01}. 
Our derived abundance ratios for one star in the spectroscopic sample, S58, are much higher than the observed values. The reason for this difference
is not known.

\item 
There are significant differences in the SFH as a function of radius.
Star formation was more vigorous and longer lived in the central
region than in the outer region, resulting in a metallicity gradient.
We conclude that the primary driver for the radial stellar population gradient 
in this galaxy is the SFH.

\end{enumerate}

Recently there are several works showing the importance of outflow
(via galactic wind or via external mechanisms like ram pressure stripping)
in the chemical evolution of dSphs \citep{lan04, lan07, rec08, mat08}.
We are planning to investigate the effect of outflow in the chemical evolution
of dSphs using the SMART model in the subsequent paper.

\acknowledgments

M.G.L. is supported in part by the grant from the Basic Research Program from of the Korea Science \& Engineering Foundation (R01-2007-000-20336-0). 
Numerical simulations were performed using
the Linux Cluster for Astronomical Computations of KASI-ARCSEC.
DZ acknowledges support from NSF AST-0307482 and NASA LTSA 04-000-0041. DZ
gratefully acknowledges financial support during his sabbatical from the 
John Simon Guggenheim foundation, KTIP through its support from the NSF grant
PHY99-07949, and the NYU Physics Department and Center for Cosmology and Particle Physics.

\appendix

\clearpage





\begin{deluxetable}{cccccccccc}
\tablecaption{Properties of the regions in the Sextans dSph
\label{region_def_tbl}}
\tablewidth{0pt}
\tablehead{
\colhead{Region\tablenotemark{a}} &
\colhead{Radius\tablenotemark{b}} &
\colhead{Area\tablenotemark{c}} &
\colhead{$N_1$\tablenotemark{d}} &
\colhead{$N_2$\tablenotemark{e}} &
\colhead{$\rho_1$\tablenotemark{f}} &
\colhead{$\rho_2$\tablenotemark{g}}
}
\startdata
R1  & $R < 7.\!'3$       & 167.4 & 4318 & 4259 & 25.8 & 25.4 \\
R2  & $7.\!'3 < R < 11'$ & 212.7 & 4697 & 4633 & 22.1 & 21.8 \\
R3  & $11' < R < 15'$    & 312.3 & 5431 & 5363 & 17.4 & 17.2 \\
R4  & $15' < R $         & 483.6 & 5864 & 5795 & 12.1 & 12.0 \\
\enddata
\tablenotetext{a}{The name of each region.}
\tablenotetext{b}{The boundary of each region in arcmin.}
\tablenotetext{c}{The area of each region in arcmin$^2$.}
\tablenotetext{d}{Total number of stars in each region.}
\tablenotetext{e}{The number of stars excluding the blue straggler stars in the pentagon shown in Figure~\ref{cmd_region}.}
\tablenotetext{f}{Number density of stars in each region ($= N_1$ / Area).}
\tablenotetext{g}{Number density of stars without blue stragglers in each region ($= N_2$ / Area).}
\end{deluxetable}

\begin{deluxetable}{ccccccccccccc}
\tablecaption{Stars in the Sextans dSph with spectroscopic data by \citet{she01}
\label{spec_star}}
\rotate
\tabletypesize{\scriptsize}
\tablewidth{0pt}
\tablehead{
\colhead{Star ID\tablenotemark{a}} &
\colhead{R.A.\tablenotemark{b}} &
\colhead{Decl.\tablenotemark{c}} &
\colhead{$V$\tablenotemark{d}} &
\colhead{$B-V$\tablenotemark{e}}  &
\colhead{$V-I$\tablenotemark{f}}  &
\colhead{Age\tablenotemark{g}} &
\colhead{[Fe/H]\tablenotemark{h}} &
\colhead{[$\alpha$/Fe]\tablenotemark{i}}
}
\startdata
S35 & $\rm{10^{h} 13^{m} 57.\!^{s}72}$ & $-1^{\circ} 38' 53''.0$ & 17.30 & 1.45 & 1.37 & 15 & $-1.93 \pm 0.11$ & $0.13 \pm 0.06$ \\
S56 & $\rm{10^{h} 12^{m} 41.\!^{s}86}$ & $-1^{\circ} 45' 27''.0$ & 17.33 & 1.43 & 1.39 & 15 & $-1.93 \pm 0.11$ & $0.12 \pm 0.07$ \\
S49 & $\rm{10^{h} 13^{m} 11.\!^{s}57}$ & $-1^{\circ} 43' 01''.4$ & 17.55 & 1.16 & 1.24 & 15 & $-2.85 \pm 0.13$ & $-0.01 \pm 0.11$ \\
S58 & $\rm{10^{h} 12^{m} 16.\!^{s}13}$ & $-1^{\circ} 45' 47''.9$ & 17.65 & 1.27 & 1.29 & 6  & $-1.45 \pm 0.12$ & $-0.26 \pm 0.07$ \\
S36 & $\rm{10^{h} 13^{m} 04.\!^{s}96}$ & $-1^{\circ} 39' 13''.4$ & 17.94 & 1.16 & 1.18 & 15 & $-2.19 \pm 0.12$ & $0.10 \pm 0.07$ \\
\enddata
\tablenotetext{a}{ID given by \citet{sun93}. Corresponding IDs in \citet{lee03}  are 162, 169, 195, 212, and 255.}
\tablenotetext{b}{R. A. (2000) given by \citet{lee03}.}
\tablenotetext{c}{Dec. (2000)  given by \citet{lee03}.}
\tablenotetext{d}{$V$ magnitude given by \citet{lee03}.}
\tablenotetext{e}{$B-V$ color given by \citet{lee03}.}
\tablenotetext{f}{$V-I$ color given by \citet{lee03}.}
\tablenotetext{g}{Age of star in Gyr estimated by \citet{tol03}
using the Yale-Yonsei isochrones \citep{yi01}. New estimate for S58 is
7 Gyr based on the revised Yale-Yonsei isochrones \citep{dem04}. }
\tablenotetext{h}{[Fe/H] given by \citet{tol03}.}
\tablenotetext{i}{[$\alpha$/Fe] given by \citet{tol03}.}
\end{deluxetable}


\begin{deluxetable}{ll}
\tablecaption{Model parameters
\label{model_par}}
\tablewidth{0pt}
\tablehead{
\colhead{item} &
\colhead{value}
}
\startdata
distance modulus $(m-M)_0$\tablenotemark{a} & 19.67 \\
reddening $E(B-V)$\tablenotemark{b}         & 0.03  \\
IMF index\tablenotemark{c}                & --1.35 \\
mass range                     & $0.08 M_{\odot} - 40 M_{\odot}$ \\
chemical evolution model                   & closed box model \\
initial composition\tablenotemark{d}      & primordial value\\
\enddata
\tablenotetext{a}{Determined in this study.} 
\tablenotetext{b}{Determined in this study.} 
\tablenotetext{c}{Index of the Salpeter's IMF ($x$) \citep{sal55}.}
\tablenotetext{d}{Homogeneous big bang composition \citep{wal91}.}
\end{deluxetable}

\clearpage

\begin{deluxetable}{ccccccc}
\tablecaption{Star formation history for Case A
\label{sfh_w_bs_tbl}}
\tablewidth{0pt}
\tablehead{
\colhead{Age (Gyr)\tablenotemark{a}} &
\colhead{SFR (R1)\tablenotemark{b}} &
\colhead{SFR (R2)\tablenotemark{c}} &
\colhead{SFR (R3)\tablenotemark{d}} &
\colhead{SFR (R4)\tablenotemark{e}}
}
\startdata
  0 --  4 & $0.00 \pm 0.01$ & $0.00 \pm 0.01$ & $0.00 \pm 0.01$ & $0.00 \pm 0.01$ \\
  4 --  7 & $0.03 \pm 0.01$ & $0.02 \pm 0.01$ & $0.01 \pm 0.01$ & $0.01 \pm 0.01$ \\
  7 --  9 & $0.10 \pm 0.05$ & $0.07 \pm 0.04$ & $0.06 \pm 0.03$ & $0.04 \pm 0.01$ \\
  9 -- 11 & $0.19 \pm 0.17$ & $0.16 \pm 0.12$ & $0.14 \pm 0.08$ & $0.07 \pm 0.04$ \\
 11 -- 13 & $0.61 \pm 0.26$ & $0.53 \pm 0.19$ & $0.30 \pm 0.13$ & $0.13 \pm 0.07$ \\
 13 -- 14 & $0.99 \pm 0.34$ & $0.86 \pm 0.25$ & $0.78 \pm 0.18$ & $0.64 \pm 0.11$ \\
 14 -- 15 & $0.99 \pm 0.14$ & $0.87 \pm 0.12$ & $0.78 \pm 0.10$ & $0.66 \pm 0.06$ \\
\enddata
\tablenotetext{a}{Age bin in Gyr.}
\tablenotetext{b,c,d,e}{The star formation rates of the central region R1 ($R < 7.\!'3$),
the inner region R2 ($7.\!'3 < R < 11'$), the intermediate region R3 ($11' < R < 15'$),
and the outer region R4 ($15' < R $), respectively.
These were normalized with respect to  the lifetime 
average star formation rate for the entire observed region of Sextans, 
$1.7 \times 10^{-4}$ $M_\odot$ yr$^{-1}$. 
}
\end{deluxetable}

\begin{deluxetable}{ccccccccc}
\tablecaption{Star formation history for Case B
\label{sfh_wo_bs_tbl}}
\tablewidth{0pt}
\tablehead{
\colhead{Age (Gyr)\tablenotemark{a}} &
\colhead{SFR (R1)\tablenotemark{b}} &
\colhead{SFR (R2)\tablenotemark{c}} &
\colhead{SFR (R3)\tablenotemark{d}} &
\colhead{SFR (R4)\tablenotemark{e}}
}
\startdata
  0 --  4 & $0.00 \pm 0.01$ & $0.00 \pm 0.01$ & $0.00 \pm 0.01$ & $0.00 \pm 0.01$ \\
  4 --  7 & $0.00 \pm 0.01$ & $0.00 \pm 0.01$ & $0.00 \pm 0.01$ & $0.00 \pm 0.01$ \\
  7 --  9 & $0.05 \pm 0.02$ & $0.04 \pm 0.01$ & $0.02 \pm 0.01$ & $0.02 \pm 0.01$ \\
  9 -- 11 & $0.13 \pm 0.09$ & $0.12 \pm 0.06$ & $0.08 \pm 0.04$ & $0.05 \pm 0.02$ \\
 11 -- 13 & $0.65 \pm 0.19$ & $0.41 \pm 0.14$ & $0.29 \pm 0.09$ & $0.10 \pm 0.04$ \\
 13 -- 14 & $1.07 \pm 0.29$ & $1.06 \pm 0.22$ & $0.88 \pm 0.15$ & $0.68 \pm 0.09$ \\
 14 -- 15 & $1.07 \pm 0.11$ & $1.06 \pm 0.10$ & $0.89 \pm 0.08$ & $0.77 \pm 0.06$ \\
\enddata
\tablenotetext{a}{Age bin in Gyr.}
\tablenotetext{b,c,d}{The star formation rates of the central region R1 ($R < 7.\!'3$),
the inner region R2 ($7.\!'3 < R < 11'$), the intermediate region R3 ($11' < R < 15'$),
and the outer region R4 ($15' < R $), respectively. 
These were normalized with respect to  the lifetime 
average star formation rate for the entire observed region of Sextans, 
$1.7 \times 10^{-4}$ $M_\odot$ yr$^{-1}$.}
\end{deluxetable}

\clearpage

\begin{deluxetable}{cccccc}
\tablecaption{Mean metallicity of each region in the Sextans dSph
\label{feh_mean_tbl}}
\tablewidth{0pt}
\tablehead{
\colhead{Region} &
\colhead{$\rm{<Fe/H>}$ (Case A)} & 
\colhead{$\rm{<Fe/H>}$ (Case B)} 
}
\startdata
R1 & --1.69 & --1.68 \\
R2 & --1.76 & --1.73 \\
R3 & --1.86 & --1.83 \\
R4 & --1.99 & --1.97 \\
Entire field & --1.78 & --1.76 \\
\enddata
\end{deluxetable}

\clearpage



\begin{figure}  
\epsscale{.80}
\plotone{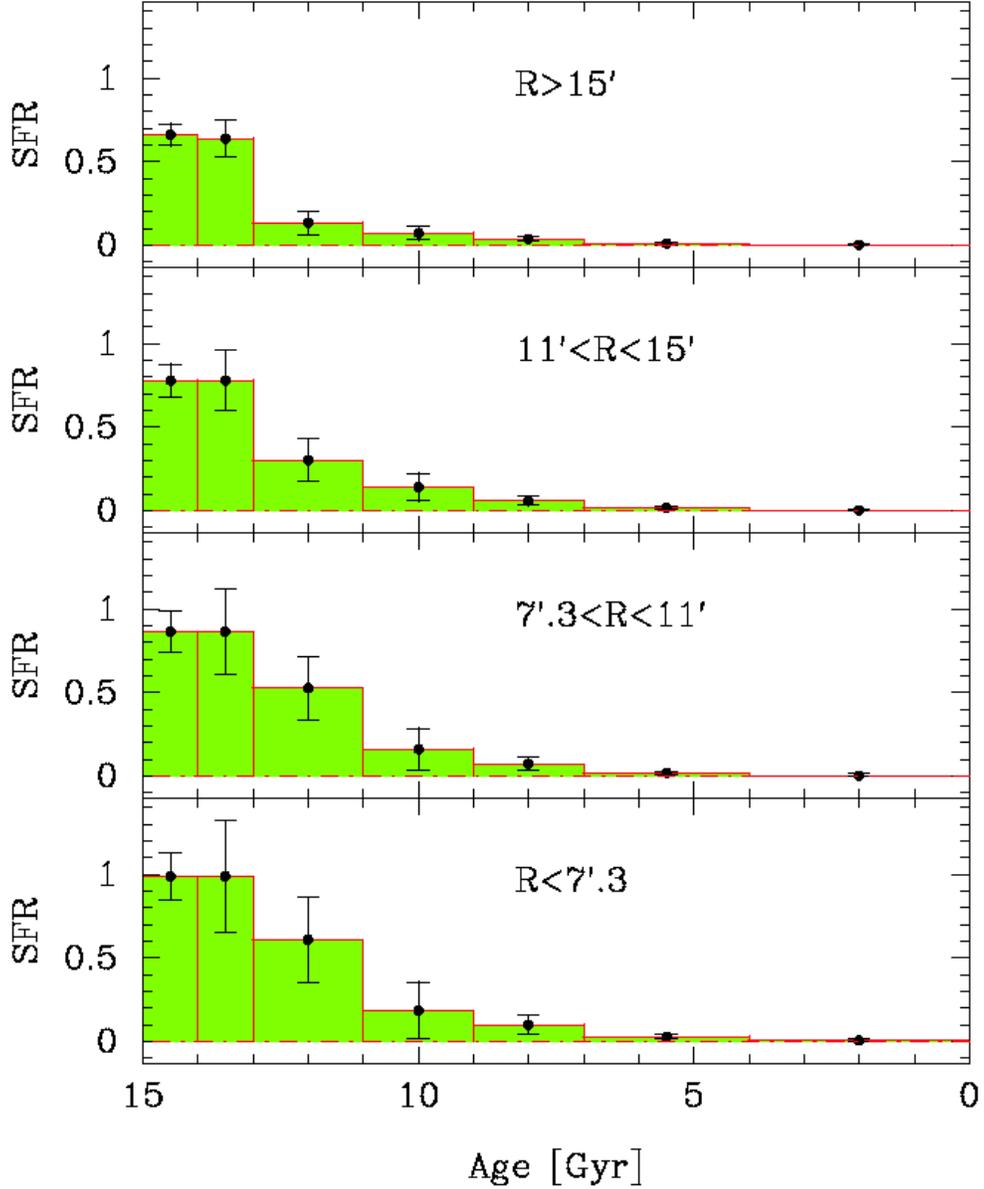} 
\caption
{Star formation history of the four regions in the Sextans dSph for Case A where blue straggler
stars were considered as main-sequence stars with intermediate age for modeling. 
The star formation rates were normalized with respect to  the lifetime 
average star formation rate for the entire observed region of Sextans, 
$1.7 \times 10^{-4}$ $M_\odot$ yr$^{-1}$.
The vertical bars indicate the asymmetric errors of 1 $\sigma$ level.
Note that a primary star formation continued for long until 7 Gyr ago, followed by
a minor star formation from 7 to 4 Gyr ago 
\label{sfh_w_bs}}
\end{figure}

\begin{figure}  
\epsscale{.80}
\plotone{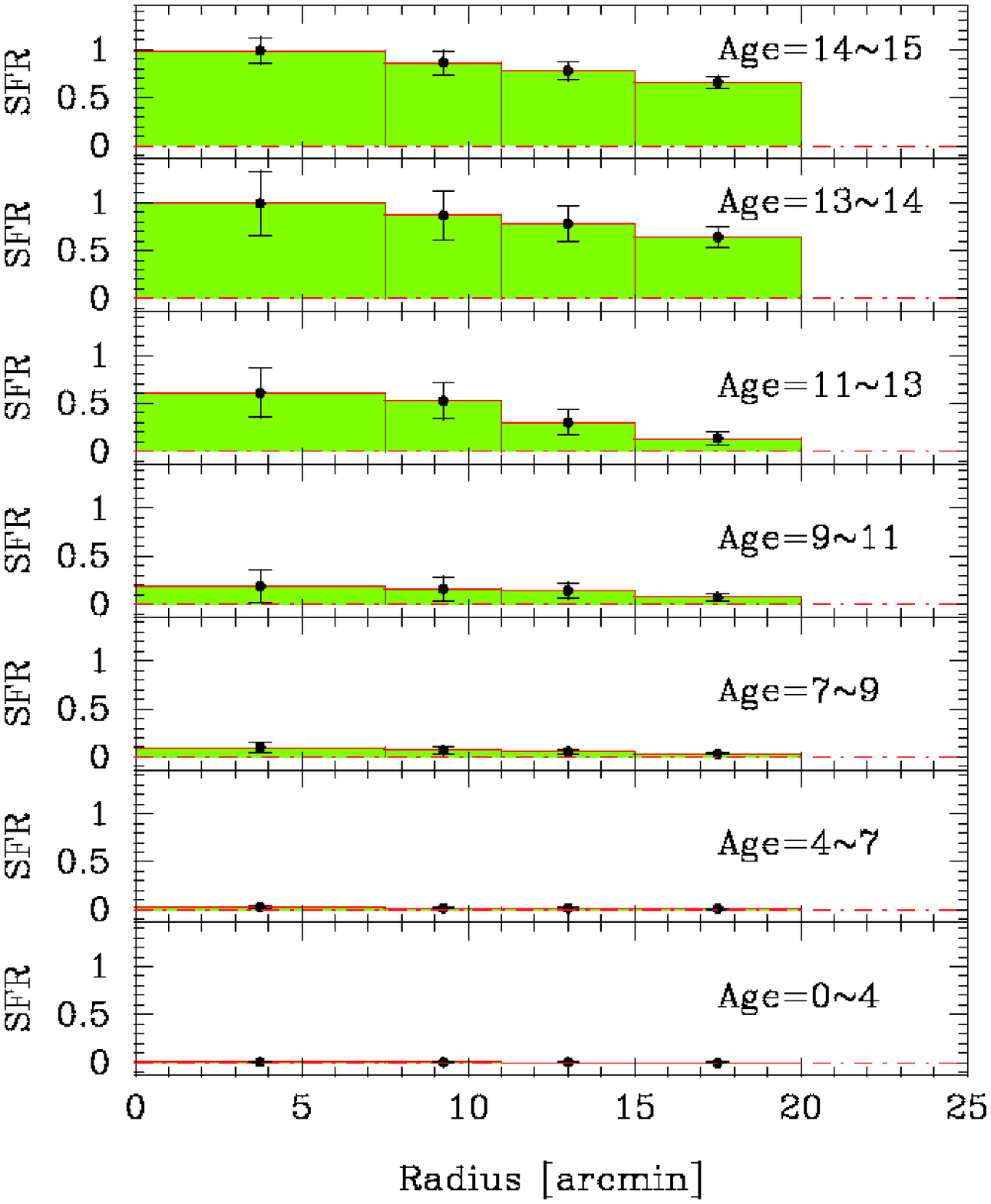} 
\caption
{Radial distribution of the star formation rates of the Sextans dSph for Case A.
The star formation rates were normalized with respect to  the lifetime 
average star formation rate for the entire observed region of Sextans, 
$1.7 \times 10^{-4}$ $M_\odot$ yr$^{-1}$.
The vertical bars indicate the symmetric errors of 1 $\sigma$ level.
Note the radial gradient of the star formation rate for extended period of time.
\label{sfh_rad_w_bs}}
\end{figure}

\begin{figure}  
\epsscale{.80}
\plotone{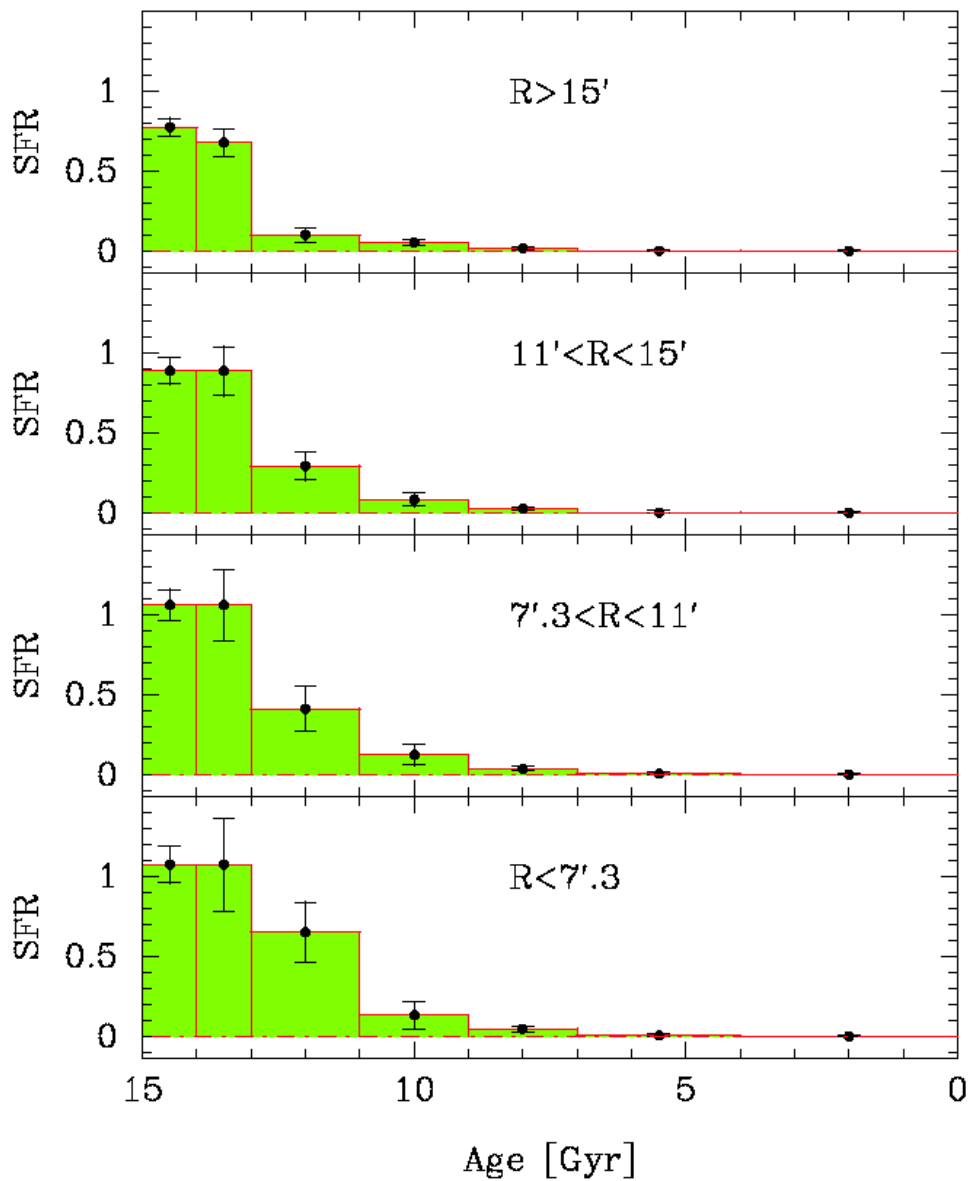} 
\caption
{Star formation history of the Sextans dSph for Case B where blue straggler
stars were not considered  for modeling.
The star formation rates were normalized with respect to  the lifetime 
average star formation rate for the entire observed region of Sextans, 
$1.7 \times 10^{-4}$ $M_\odot$ yr$^{-1}$.
The vertical bars indicate the asymmetric errors of 1 $\sigma$ level.
Note that a primary star formation continued for long until 7 Gyr ago.
\label{sfh_wo_bs}}
\end{figure}

\begin{figure}  
\epsscale{.80}
\plotone{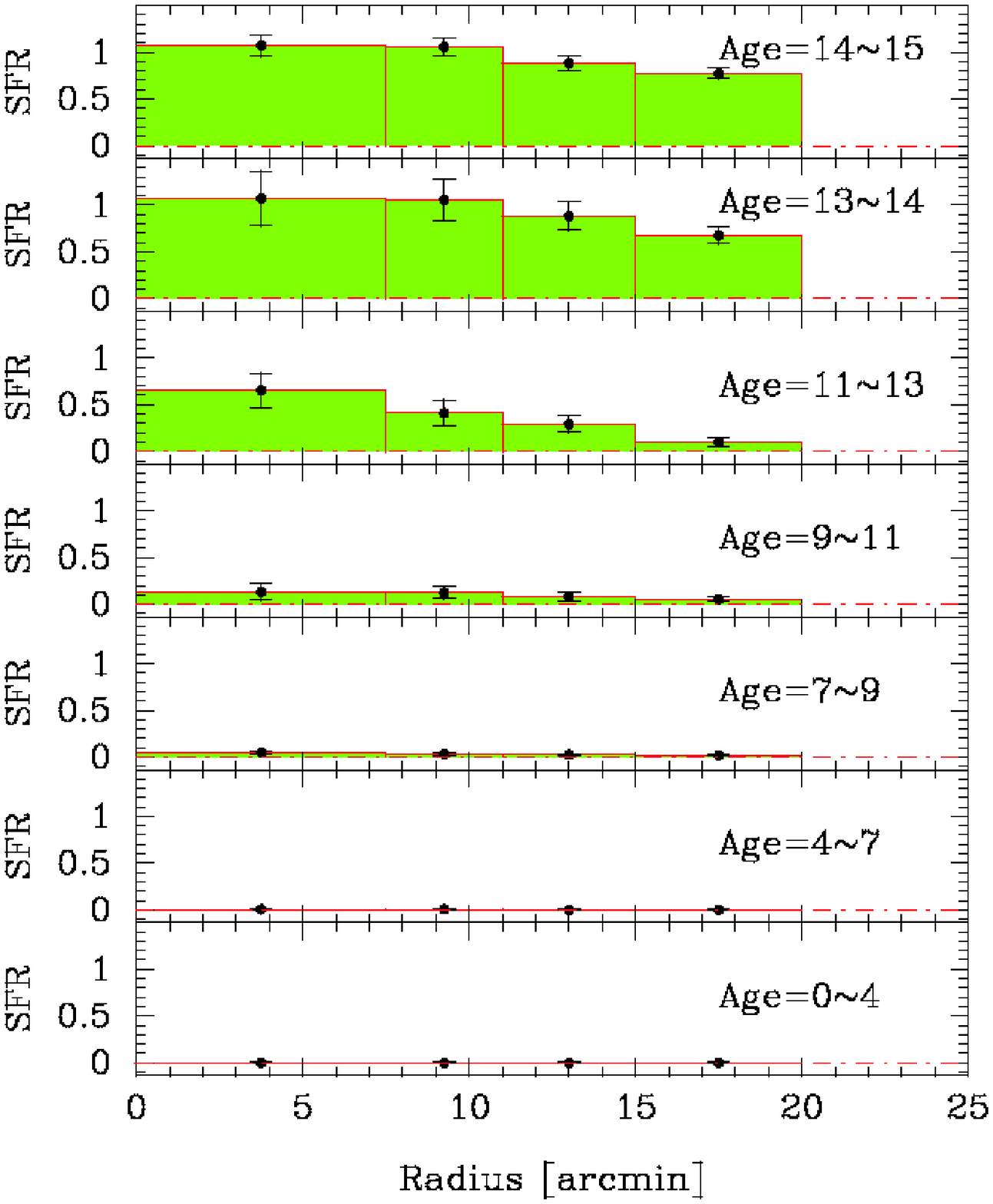} 
\caption
{Radial variation of the star formation rates of the Sextans dSph for Case B.
The star formation rates were normalized with respect to  the lifetime 
average star formation rate for the entire observed region of Sextans, 
$1.7 \times 10^{-4}$ $M_\odot$ yr$^{-1}$.
The vertical bars indicate the asymmetric errors of 1 $\sigma$ level.
Note the radial gradient of the star formation rate for extended period of time.
\label{sfh_rad_wo_bs}}
\end{figure}

\clearpage

\begin{figure}  
\epsscale{.80}
\plotone{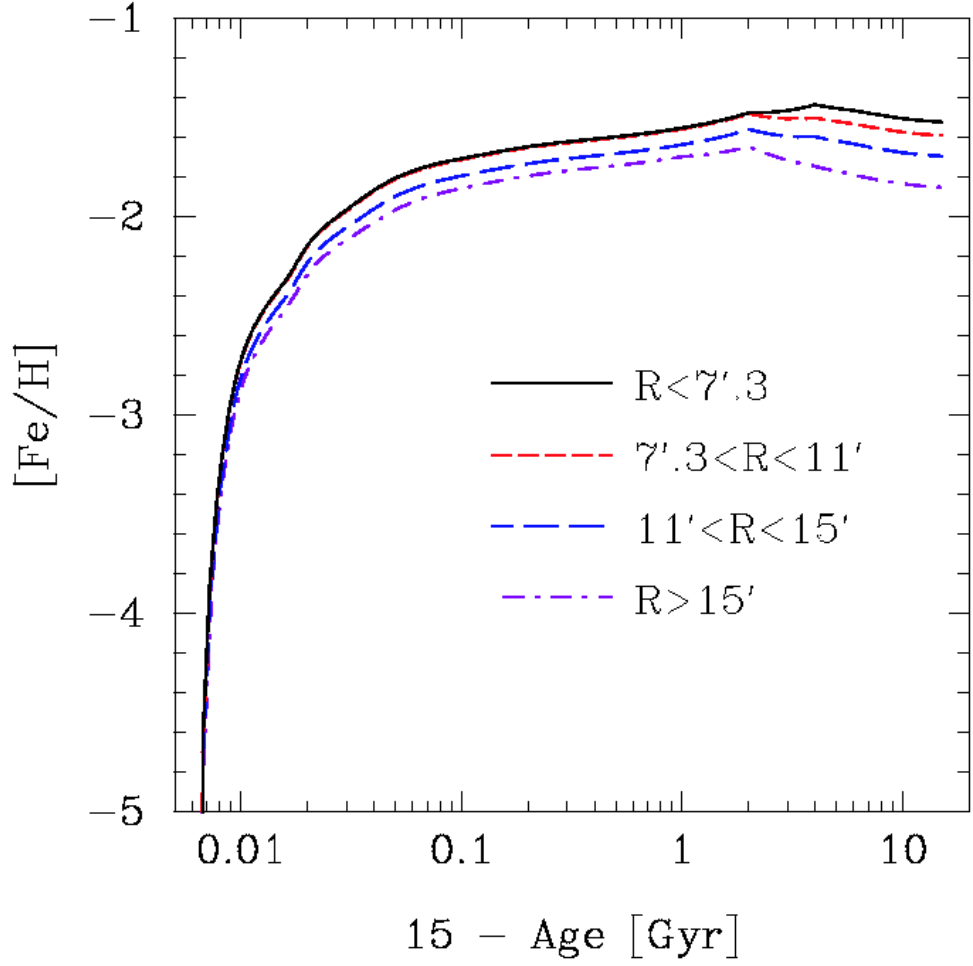} 
\caption
{Metal enrichment history for the four regions in the Sextans dSph for Case B.
The higher the lines are, the closer the corresponding regions are to the galaxy center.
(solid line for R1, short-dashed line for R2, long-dashed line for R3, and dot-dashed line for R4).
\label{feh_wo_bs}}
\end{figure}

\begin{figure}  
\epsscale{.80}
\plotone{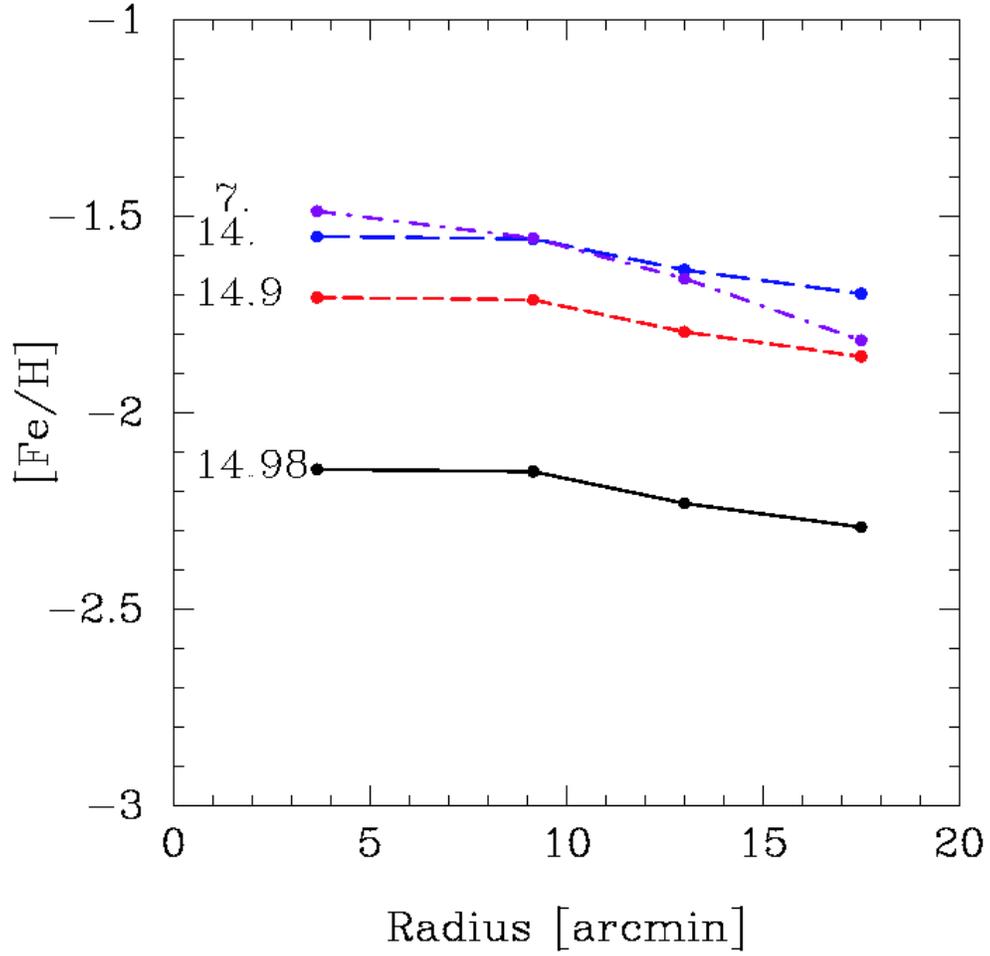} 
\caption
{Radial variation of the metallicity [Fe/H] of the Sextans dSph for Case B.
The numbers beside lines represent the age of stars in Gyr.
\label{feh_rad_wo_bs}}
\end{figure}

\begin{figure}   
\epsscale{1.00}
\plotone{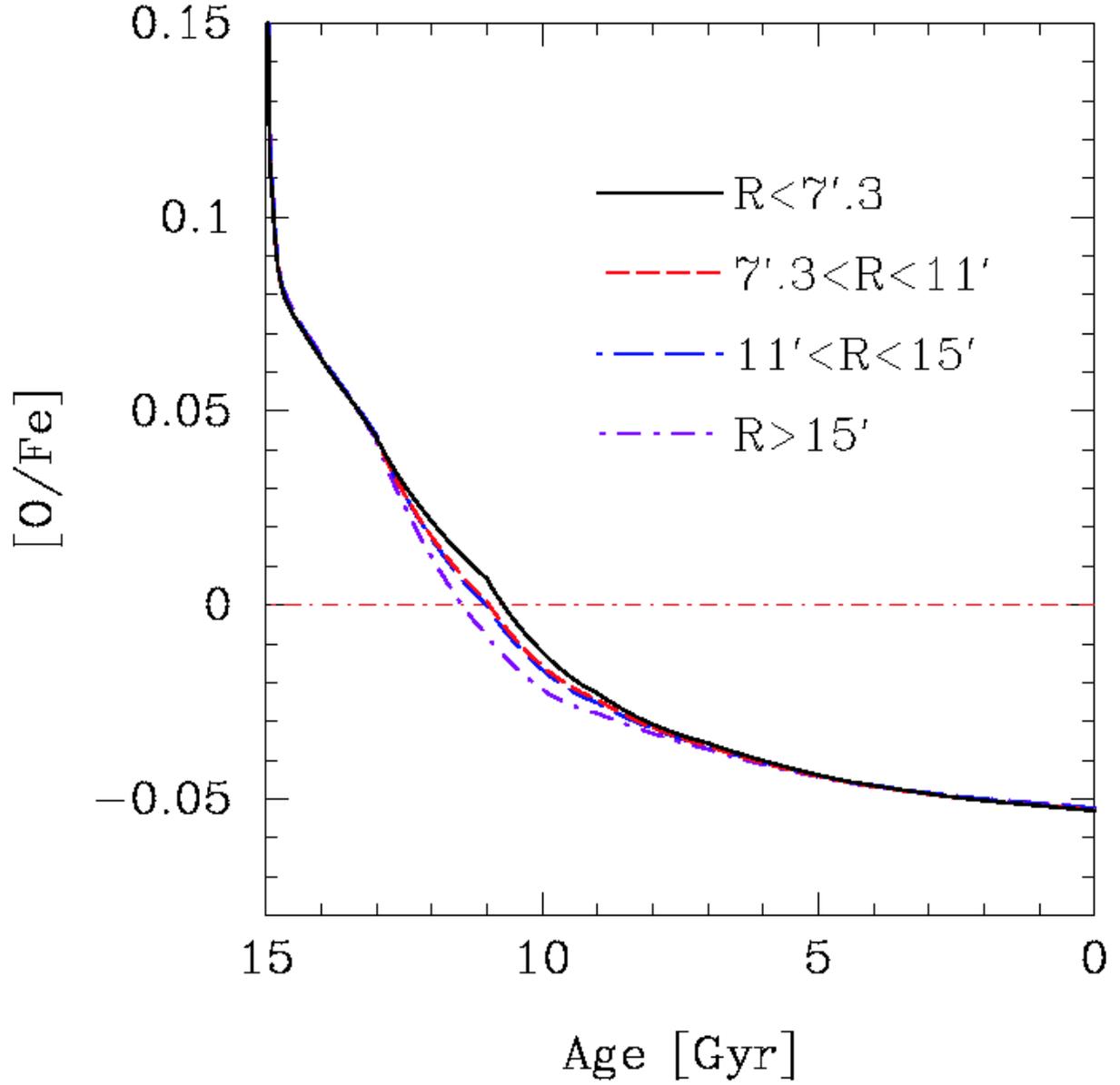} 
\caption {Evolution of [O/Fe] for the four regions in the Sextans dSph 
Case B.  \label{ofe_age_w_bs}}
\end{figure}

\begin{figure}   
\epsscale{1.00}
\plotone{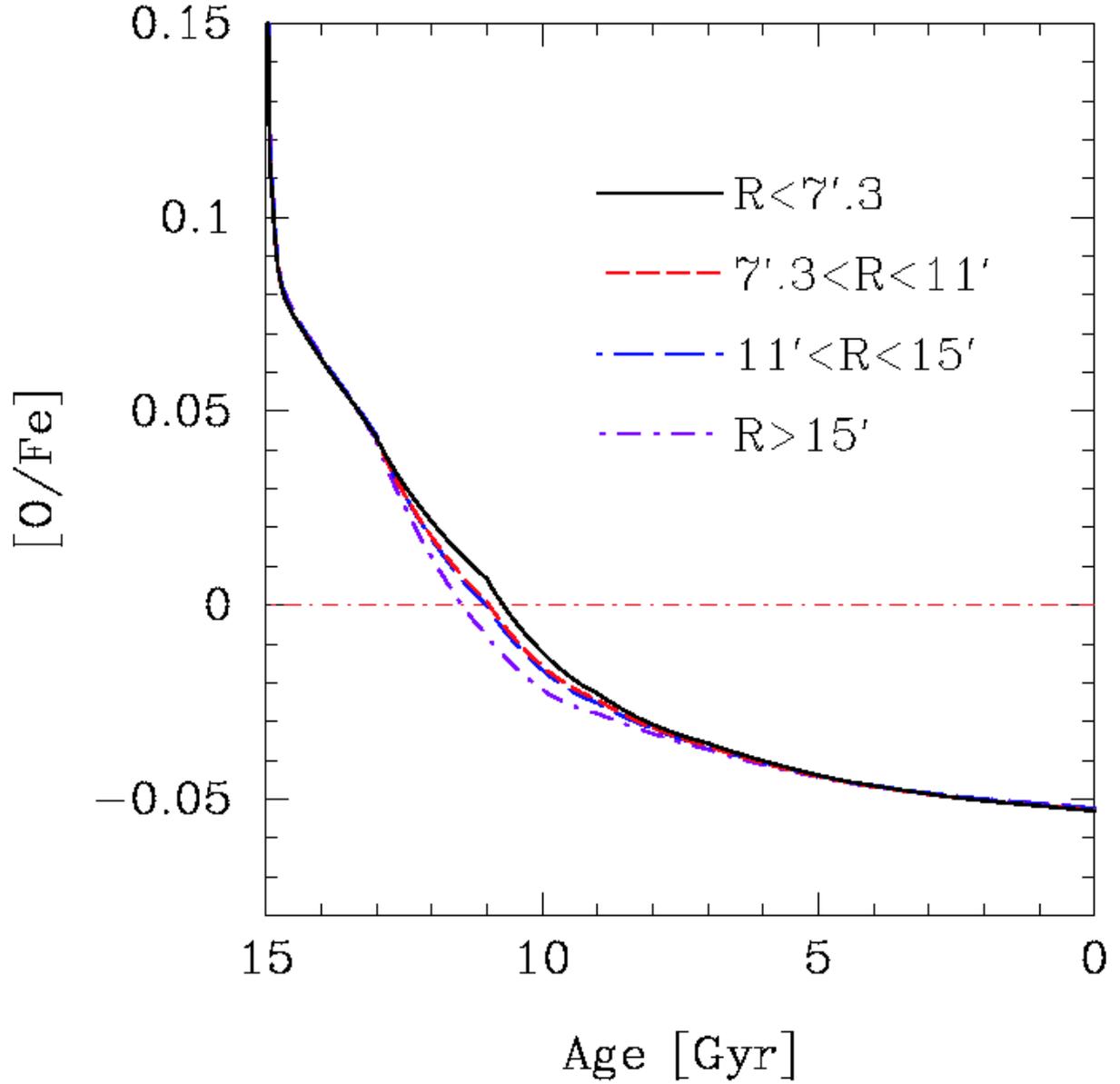} 
\caption
{Evolution of [$\alpha$/Fe] for the four regions in the Sextans dSph for Case B
derived in this study (solid lines),
where ${[\alpha/{\rm Fe}] = ( {\rm [Mg/Fe] + [Ca/Fe] + [Ti/Fe]} ) /3 }$.
Filled circles with error bars represent the observational data given by \citet{she01}
\label{alpha_age_w_bs}}
\end{figure}

\begin{figure}   
\epsscale{1.00}
\plotone{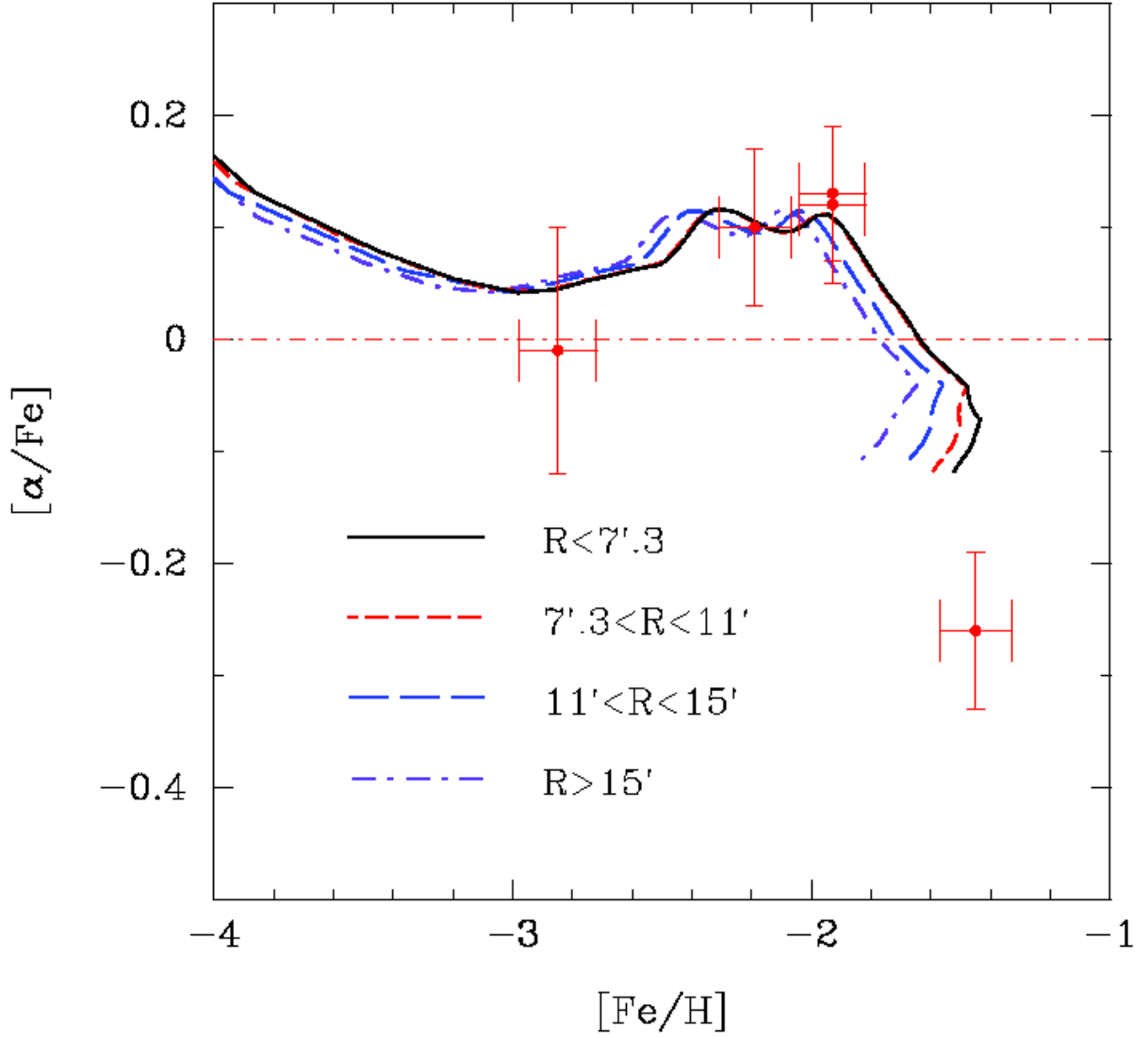} 
\caption
{[$\alpha$/Fe] ratios versus [Fe/H] for the four regions in the Sextans dSph for Case B. 
Filled circles with error bars represent the observational data given by
\citet{tol03}, and the lines represent the values derived  in this study.
\label{alpha_w_bs}}
\end{figure}

\begin{figure}   
\epsscale{1.00}
\plotone{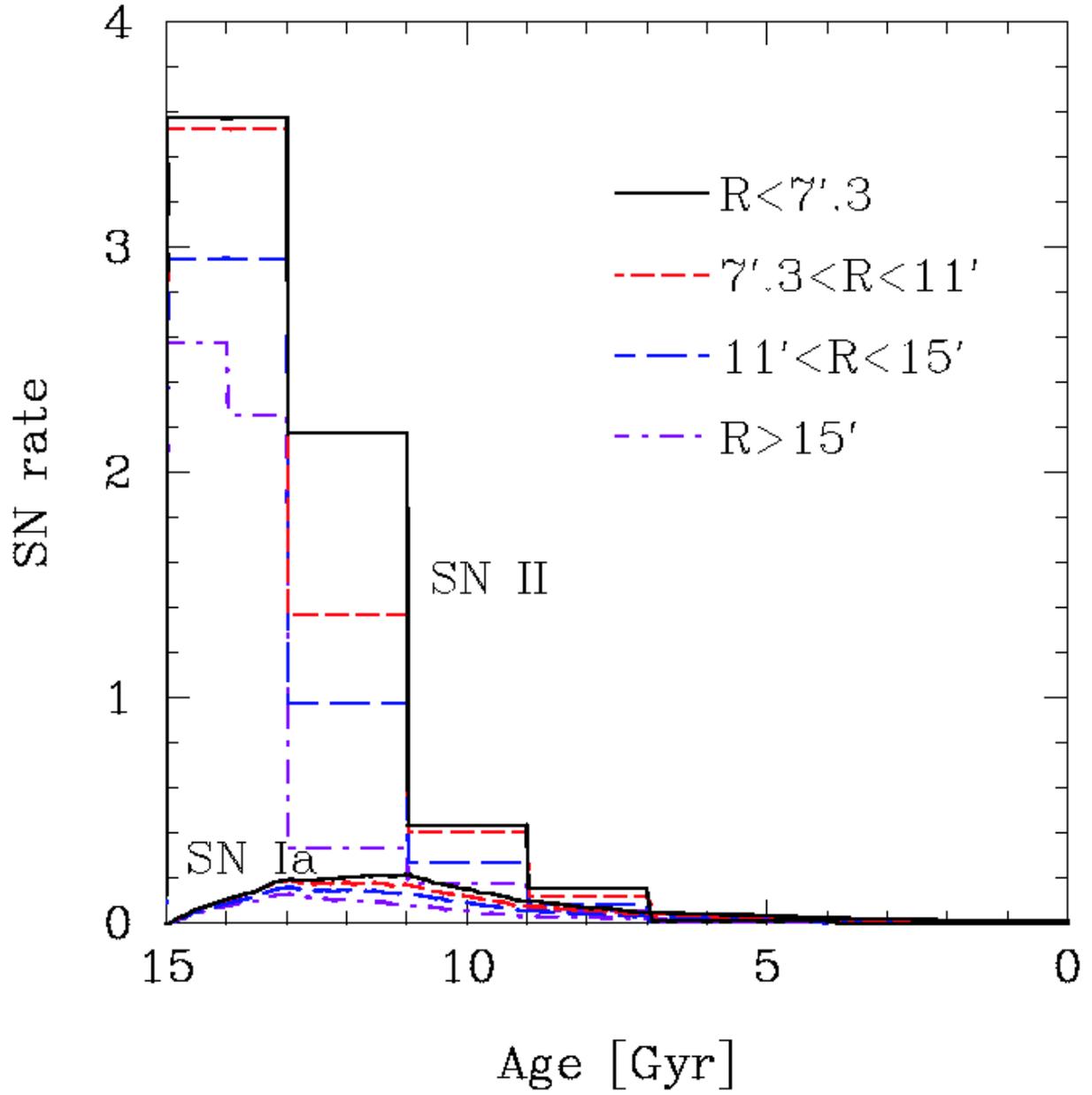} 
\caption
{Evolution of the SN Ia (lower group) and SN II (upper group)  rate for the four regions in the Sextans dSph for Case B. 
\label{snr_w_bs}}
\end{figure}

\clearpage

\begin{figure}   
\epsscale{1.0}
\plotone{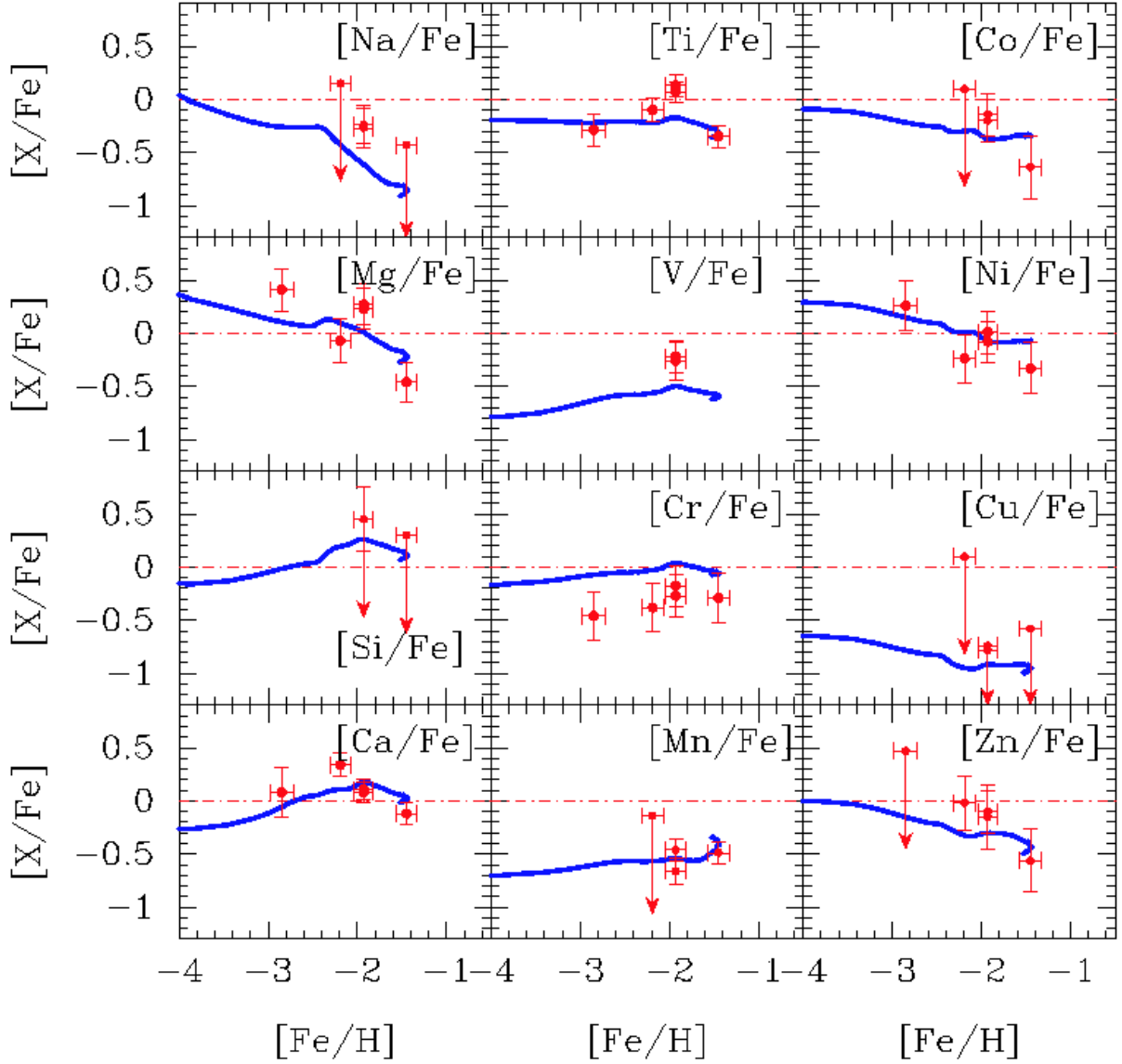} 
\caption
{Abundance ratios for several elements of the Sextans dSph
for Case B.
Filled circles with error bars represent the observational data given by
\citet{she01}, and the solid lines represent the results derived  in this study.
\label{ratio_wo_bs}}
\end{figure}

\begin{figure}   
\epsscale{.80}
\plotone{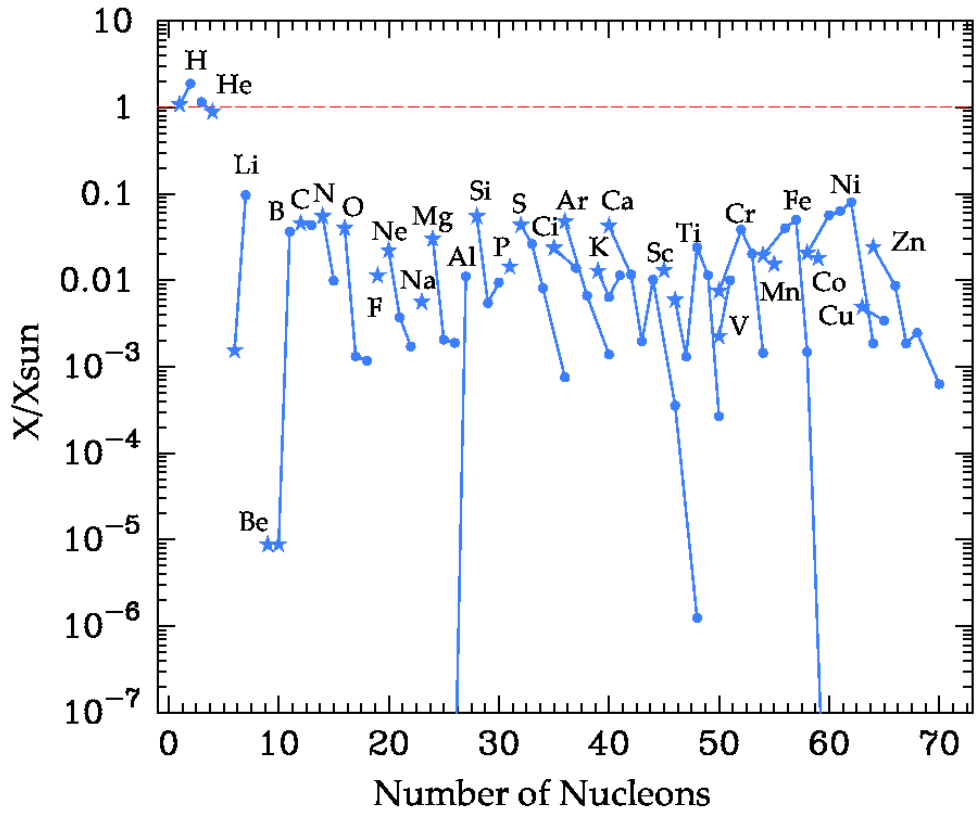} 
\caption
{The present stable isotopes of the Sextans dSph normalized to the solar abundance for Case B. The starlets represent the isotopes most abundant in the Sun.
\label{isotopes_wo_bs}}
\end{figure}
\clearpage








\end{document}